\documentclass{article}
\usepackage{graphicx} 
\usepackage{comment}

\newcommand{\ep}{\varepsilon}

\newcommand{\eq}[1]{\begin{equation} #1 \end{equation}}

\newcommand{\mtc}[1]{\mathcal{#1}}

\usepackage{amsmath}
\usepackage{amssymb}
\usepackage{fancyhdr}
\usepackage[version=4]{mhchem}

\usepackage{bm} 

\usepackage[utf8]{inputenc}   
\usepackage[T1]{fontenc} 
\usepackage{lmodern}

\title{Reformulating Chemical Equilibrium in Reacting Quantum Gas Mixtures: Particle Number Conservation, Correlations and Fluctuations}

\author{Diogo J. L. Rodrigues*}
\date{*\small Department of Physical and Analytical Chemistry, Faculty of Chemistry, University of Oviedo\\
Email: \textbf{uo302765@uniovi.es}}

\begin{document}


\maketitle

\begin{center}
\section*{Abstract}
\end{center}

\noindent The canonical‑ensemble description of reactive quantum gas mixtures is reformulated by incorporating a single global particle-number-conservation constraint over the combined spectra of inter-converting species. This constraint replaces the conventional equality of chemical potentials. Fermi–Dirac or Bose–Einstein correlations naturally emerge across one-particle energy eigenstates of species sharing identical spin-statistics, which in ergodic single-systems manifest as intrinsic features of the equilibrium state. By embedding all microstates linked by conversion pathways, the framework incorporates concentration fluctuations in the statistical description. The formalism offers fresh insights into quantum chemical equilibrium in reactive mixtures with composition fluctuations and smoothly reduces to the classical ideal gas limit via an extended partition function that generalizes classical chemical‑equilibrium treatments.\\

\noindent \textbf{Keywords}: Statistical mechanics, Quantum statistics, Canonical ensemble, Partition function, Chemical potential, Concentration fluctuations

\maketitle

\newpage

\section{Introduction}
\label{intro} 

Chemical equilibrium in many-body systems is a widely researched topic, spanning scales from that of particle physics \cite{Cleymans2010,Tawfik2014} to large biological macromolecules \cite{Mclean1938}. Usual treatments of chemical equilibrium, wether involving chemical reactions or simpler forms of mixture stability, generally consist of classical or semiclassical approaches \cite{Koudriavtsev2012-hc,Lawrence1999, Lehmann1996,Zimmermann1985,Gau1981}. Nevertheless, in some instances and under different circumstances, classical treatments fall short due to emerging quantum effects \cite{Borshchevskii2025, Brogioli2013, levy2018quantum}. This breakdown challenges the limits of classical, and of many semiclassical treatments. Tentatively employed full quantum treatments of (reacting) mixture stability are usually dynamical in nature or based on scattering theory \cite{Yukalov2022, Zhang20231466, Jachymski2022, He2020, Croft2020, Idziaszek2010, Julienne2009,  Lawrence1999, Stolz1979}. These are of a much more complex nature, and formal solutions are practical only for the simplest systems or are obtained under important approximations.
Reactions at the quantum level have also become increasingly important for trending research topic such as ultracold physics and chemistry \cite{Karman2024722, Zhang20231466, Zhao2022, Zhang2021708, Liu2020, Hu2019, Quéméner20124949, Ospelkaus2010853, Carr2009, Bell2009}.

In this work we contribute to this state-of-the-art by introducing a novel and purely canonical treatment for mixtures of interconverting quantum gases, of which chemical reactions are a special case. The framework naturally incorporates composition fluctuations arising from overall particle number conservation and describes systems of reacting molecules or atoms where quantum effects are relevant and quantum statistics must be considered.
We aim to describe a reacting mixture of quantum gases where each individual particle is described by the canonical Gibbs state \cite{hillthermo}.
The approach employed is based on the maximum entropy principle and makes use of Lagrange multipliers to formulate a composite system characterized by a single chemical potential that is associated with a single constraint of particle conservation over the entirety of the composite system.

For clarity and tractability, we restrict our model to weakly interacting or thermalized non-interacting quantum gases and start by focusing in describing one-to-one reaction schemes of (true or apparent) unimolecular nature \cite{Quack1981}.
Reactions strictly of this type fall in the category of isomerization and rearrangement reactions (which can involve bond-breaking and bond-formation). 
The replacement of one-to-one reactions to one reactant species to one product species, but with different reaction coefficients is further introduced as a simple generalization. This allows for handling bond-formation or bond-breaking between different molecules in a single reaction. 
Though we do not discuss it, the treatment should be generalizable to any type of complex reaction networks, in an unrestricted form.

Rather than focus on the specific details involved in describing real systems, we aim to build intuition regarding chemical equilibrium of interconverting/chemically reacting quantum gases, simultaneously incorporating composition fluctuations via distinct microstates, improving over traditional methods.
We also delve in the classical gas limit of our approach, gathering insights into how to build improved classical partition functions that incorporate fluctuating behavior and quantum effects.

This work is organized as follows. In Section \ref{sec1}, we outline the traditional derivation of quantum distributions for independent subsystems using the method of Lagrange multipliers. Section \ref{sec2} extends this formalism to the case of chemically reacting/interconverting species under a unified chemical potential, with the necessary approximations and assumptions highlighted and discussed. In Section \ref{sec3}, we move our framework to the classical limit, 
and it is shown, using analysis of variances, how our formulation improves on traditional approaches to equilibrium of classical gases.
Finally, a few concluding remarks follow.


\section{Review of quantum distributions for independent subsystems} 
\label{sec1} 

We start by reviewing, while establishing our notation, the statistical description for independent subsystems of weakly-interacting quantum gases. This approach is also valid for gases that are non-interacting within themselves - provided they are sufficiently thermalized by interacting with an environment. 
The number of available microstates for a canonical ensemble of indistinguishable (approximately non-interacting) bosons or fermions is given, respectively, by:
\eq{
\label{eq1}
W^{BE}=\prod_{s=1}^{\mathcal{S}} \frac{(g_s+n_s -1)!}{(g_s -1)!n_s!} \quad \quad W^{FD}=\prod_{s=1}^{\mathcal{S}} \frac{g_s!}{(g_s - n_s)!n_s!}
}
where $BE$ and $FD$ denote the \textit{Bose-Einstein} and \textit{Fermi-Dirac} distributions. Here, $g_s$ is the degeneracy of the energy level labeled by $s$, with average occupation $n_s$. $\mathcal{S}$ is the total number of available energy levels. \footnote{The factorial notation $(...)!$ is generalized for non-integer values using the gamma function \cite{artin2015gamma}.} \footnote{From here on the $s=1$ limit in the sum will be replaced simply by $s$.} 
To derive the BE and FD distributions, one resorts to the method of Lagrange multipliers, variationally maximizing the number of microstates in Equation (\ref{eq1}) with respect to the occupations $n_s$, while imposing the following constraints on the system:  
\begin{itemize}
\item $ \sum_s^{\mathcal{S}} n_s \varepsilon_s = E$, meaning energy conservation around a well-defined average energy $E$; \footnote{For molecules, the energy of a level $s$ must also specify the quantum numbers for its internal structure. If we label by a single $s$ all the possible combinations of translation (for free molecules) and internal degrees of freedom (dof) (generally rotational, vibrational and electronic degrees of freedom), we denote: 
$\ep_s \equiv \sum_{dof} \ep_{dof, s} = \ep_{trs,s_t} + \ep_{rot,s_r} + \ep_{vib,s_v} + \ep_{elec,s_e}$. Thus, each $s$ is used to specify one set of possible quantum numbers 
$(s_t,s_r,s_v,s_e)$. 
Likewise, $g_s$ is then the product of the degeneracy of every degree of freedom $g_s \equiv \prod_{dof} g_{dof,s} = g_{s_t}g_{s_r}g_{s_v}g_{s_e}$.
This is of course generalizable for any and all potential degrees of freedom.}
\item $\sum_s^{\mathcal{S}} n_s = N $, which imposes conservation of particle number $N$. 
\end{itemize}
One valid choice of variational Lagrangian function relies on maximizing $\ln{W}$, which shares the same maxima of $W$:
\begin{equation}
\label{eq2}
\mathcal{L}= \ln{W} - \beta \left( \sum_s^{\mathcal{S}} n_s \varepsilon_s - E \right) + \beta \mu \left( \sum_s^{\mathcal{S}} n_s - N \right)
\end{equation}  
where the Lagrange multipliers have already been substituted by their known expressions. $\beta = 1/k_B T$ is the inverse temperature, and $\mu$ is the chemical potential. Using Stirling's approximation for the factorial terms, at the $g_s, n_s \gg 1$, with $n_s/g_s \sim \mathcal{O}(1)$ limit one obtains the following form of the occupations \footnote{It is worth mentioning that Equation (\ref{eq3}) is the solution one obtains using the Darwin-Fowler method, without the need to resort to the specified approximations $g_s, n_s \gg 1$ and $n_s/g_s \sim \mathcal{O}(1)$ \cite{darwin1922xliv}. This makes the derivation more general than it might seem at a first glance.}:
\begin{equation}
\label{eq3}
n_s = \frac{g_s}{e^{\beta \varepsilon_s - \beta \mu} \mp 1},
\end{equation}  
where the $-$ sign corresponds to bosons (BE distribution) and the $+$ sign corresponds to fermions (FD distribution). 
\footnote{This optimization of occupations within an average energy constraint at the canonical level fully describes the thermodynamics of the system at the thermodynamic limit, characterized by $N,V \rightarrow \infty$ and $N/V = \mathrm{constant}$. This holds because fluctuations become negligible in the thermodynamic limit \cite{pathria2017statistical,landau2013statistical}.
For finite systems, these occupations are actually average occupations.  }

Consider now two distinguishable, thermalized weakly- or non-interacting particle types, $A$ and $B$, in a common canonical ensemble. Since there are no correlations between the particles, the statistical independence theorem holds, and the multiplicity of certain pair of levels $s \in A$ and $s' \in B$ is the product of their individual multiplicities: $W_{ss'}=W_s\times W_{s'}$. The total number of microstates for the composite system is therefore given by:
%
%
\begin{align}
\label{eq4}
& W^{BE}_{A \times B} = W^{BE}_A \times W^{BE}_B \notag \\
& = \prod_{s \in A}^{\mathcal{S}_A} \frac{(g_s + n_s - 1)!}{(g_s - 1)! n_s!} \times \prod_{s \in B}^{\mathcal{S}_B} \frac{(g_s + n_s - 1)!}{(g_s - 1)! n_s!}, \notag \\ 
\\
& W^{FD}_{A \times B} = W^{FD}_A \times W^{FD}_B \notag \\
& = \prod_{s \in A}^{\mathcal{S}_A} \frac{g_s!}{(g_s - n_s)! n_s!} \times \prod_{s \in B}^{\mathcal{S}_B} \frac{g_s!}{(g_s - n_s)! n_s!} \notag
\end{align}  
where $\mathcal{S}_A$ and $\mathcal{S}_B$ are the total number of energy levels available to $A$ and $B$, respectively. Combinations of independent bosons and fermions can also be described this way using a single number of microstates $W$ consisting of bosonic and fermionic independent parts, $W=W^{BE}\times W^{FD}$.
Assuming that under thermal equilibrium with the same environment the inverse temperature $\beta$ is the same for both subsystems, but the chemical potentials can differ, maximizing the entropy of Equation (\ref{eq4}) for a generalized system of any number of independent subsystems $\{ \mtc{A} \} = \{ A, B, C, ... \}$, with microstate number $W_{\times \mtc{A}}$, is equivalent to maximizing the following Lagrangian function:
\begin{align}
\label{eq5}
\mathcal{L} = \ln{W_{\times \mtc{A}}} &- \sum_{\mathcal{A}} \beta \left( \sum_{s \in \mathcal{A}}^{\mathcal{S}_{\mathcal{A}}} n_s \varepsilon_s - E^{\mathcal{A}} \right) \\
&+ \sum_{\mathcal{A}} \beta \mu^{\mathcal{A}} \left( \sum_{s \in \mathcal{A}}^{\mathcal{S}_{\mathcal{A}}} n_s - N^{\mathcal{A}} \right) \notag
\end{align}  
$E^{\mathcal{A}}$ is the contribution to the average energy due to species $\mathcal{A}$, and $N^{\mathcal{A}}$ the number of $\mathcal{A}$ particles (which are assumed to be \textit{a priori} known input variables).
\footnote{In this work, except to avoid cluttering of equations, particle labels will be given as superscripts and the label of the subsystems as a subscript.
We will, when useful for clarity and simplicity, denote properties of levels $s\in A$ as $g_s^A$, $n_s^A$ or $\varepsilon_s^A$.}
We thus obtain, from the variational procedure, occupations of the form:
\begin{equation}
\label{eq4.1}
n^{\mathcal{A}}_s = \frac{g^{\mathcal{A}}_s}{e^{\beta\varepsilon^{\mathcal{A}}_s -\beta\mu^{\mathcal{A}}} \mp 1}
\end{equation}
i.e. each following Equation (\ref{eq3}) with their own chemical potential $\mu^{\mathcal{A}}$. We use the ${\mtc{A}}$ superscripts to remind us that we are obtaining the occupation of a level belonging to subsystem ${\mtc{A}}$.

It may happen that only some of the molecular degrees of freedom contributing to the energy eigenstates are effectively thermalized. In those situations, only these thermalized degrees of freedom should be included in the treatment, providing a equilibrium chemical potential in a restricted form.

\section{Quantum distributions for interconverting subsystems in chemical equilibrium}
\label{sec2}

Up to this point, our analysis has focused on standard statistical physics applied to weakly interacting gas systems, where quantum correlations arise exclusively among indistinguishable particles within the same species. Particles belonging to different, distinguishable species are treated as completely uncorrelated. 
We now introduce a similar framework that incorporates a simple form of inter-species correlation that preserves the weakly interacting (or non-interacting) nature of the system: conversion (for example, via chemical reactions). We call this a {\textit{conversion-coupled}} system. \footnote{The formalism herein described applies to any type of conversion between quantum species. We will sometimes specifically focus on the special case of chemical reactions, notwithstanding the generality of the treatment.}

\subsection{Variational approach and the form of microstate counting}
\label{subsec3.1}

Let's assume as a starting point one-to-one reactions $\ce{A <=> B}$, where $\ce{A}$ and $\ce{B}$ are (thermalized) species obeying the same spin-statistics. Under chemical equilibrium, the chemical potential of the species must match at all times, i.e., $\mu^A = \mu^B \equiv \mu$. 
Thus, we should be able to statistically describe the equilibrium system by imposing, in the variational formulation, a single particle number conservation constraint, associated with a multiplier which is this unique $\mu$. The associated conserved particle number is $N= N^A +  N^B = \langle N^A \rangle +  \langle N^B \rangle $, the total particle number of the system and the sum of every possible instantaneous pair of subsystem occupations which describes the system at any given instant (which also equals the sum of subsystem occupation averages). 
When generalized for a system of any number of species (which might include any number of long-lived intermediates such as $\ce{A <=> B <=> C ...}$), connected via one-to-one reactions, the constraint introduced in the Lagrangian function is the conservation of the total particle number: 
\eq{\label{eq3.?}
N= \sum_{\mathcal{A}}\langle N^{\mathcal{A}} \rangle}
Unlike in the independent subsystem scenario, the subsystem average particle numbers $\langle N^{\mtc{A}} \rangle$ will be different from that of possible initial conditions or values found outside of equilibrium, due to the presence of conversion freedom. The independent subsystem particle number, on the other hand, always remains a conserved quantity.
This freedom has particles of the system transverse the reaction pathway and interconvert.
This additional freedom in the statistical treatment (when compared to the independent subsystem approach) must also be taken into account such that it reflects itself on the average energy of the system and its conservation constraint. 
For thus, we similarly to Equation (\ref{eq3.?}) write the (average) conservation of energy in the Lagrangian function in terms solely of a (average) total system energy $E$, which is constrained to be the sum of the average energies on each subsystem:
\eq{E= \sum_{\mathcal{A}} E^{\mathcal{A}} }
Adopting these constraints, and denoting the microstate number for a system in chemical equilibrium by $W_{A|B|...} \equiv W_{|\mtc{A}}$, we get the following chemical equilibrium Lagrangian function:
\begin{align}
\label{eq6}
\mathcal{L}= \ln{W_{|\mtc{A}}} &- \beta \left( \sum_{s \in A,B,...}^{\mathcal{S}} n^{eq}_s \varepsilon_s - E  \right) \\
&+ \beta \mu \left( \sum_{s \in A,B,...}^{\mathcal{S}} n^{eq}_s - N \right) \notag
\end{align}
$\sum_{s \in A,B,...} \equiv \sum_{s \in \{\mtc{A}\}}$ sums over the joint set of levels (over all subsystem spectra), and $\mathcal{S}=\sum_{\mtc{A}} \mathcal{S}_{\mtc{A}}$, is the sum of both subsystem' spectra size. $n^{eq}_s$ are the (chemical) equilibrium occupations of each level $s$. The sum now runs over all available levels on all available subsystems.

Equation (\ref{eq6}) is the Lagrangian function which variationally provides us a proper statistical description of the reaction system.
One can note how the constraints present on this function reduce to those of a single (sub)system (Equation (\ref{eq2})) with a spectrum that is the joint set of levels of all reacting (interconverting) subsystems, and composed of $N$ indistinguishable particles and average energy $E$.
Based on this, and since we no longer refer to well-defined subsystem constraints in building the variational problem, the subsystem partition is expected to be irrelevant to the statistical-physical description of the system, as only the energy and degeneracy of each level is needed to unequivocally characterize the total spectrum.
Since we're still considering systems of non-interacting or approximately non-interacting particles (and coupled only by the number conservation constraint) it is expected that the distribution of levels should follow the regular expected BE/FD multiplicity found for independent particles. 
This translates to a total multiplicity that is the product of the (independent) multiplicities of each level of the joint spectrum $W^{BE/FD}_{|\mtc{A}}=\prod_{s \in A,B...} W^{BE/FD}_s$, with the chemical equilibrium condition encoded in the dependence of $n^{eq}_s$ with the chemical potential $\mu$.

If we take this form of $W$, in its many-subsystem generalized form of Equation (\ref{eq4}):
\begin{align}
\label{eq7}
& W^{BE}_{| \mtc{A}} = \prod_{\mtc{A}} \left[ \prod_{s \in \mathcal{A}}^{\mathcal{S}_{\mathcal{A}}} \frac{(g_s + n^{eq}_s - 1)!}{(g_s - 1)! n^{eq}_s!} \right] \quad  \notag \\
\\
& W^{FD}_{| \mtc{A}} = \prod_{\mtc{A}} \left[ \prod_{s \in \mathcal{A}}^{\mathcal{S}_{\mtc{A}}} \frac{g_s!}{(g_s - n^{eq}_s)! n^{eq}_s!} \right] \notag
\end{align}  
with the occupations in equilibrium $n^{eq}_s$, found from the variational approach using the Lagrangian function of Equation (\ref{eq6}):
\begin{equation}
\label{eq7.1}
n^{\mathcal{A}, eq}_s = \frac{g_s^{\mtc{A}}}{e^{\beta\varepsilon^{\mathcal{A}}_s - \beta\mu} \mp 1}
\end{equation}
\footnote{We keep the $\mtc{A}$ superscripts even though in this case they are redundant, as we are working with the joint spectrum, and the chemical potential is common to all levels.} 
This distribution has the following properties:
\begin{itemize}
\item The occupation numbers within each subsystem $\mtc{A}$ are correlated the same way as in the standard independent quantum distributions; 
\item 
Correlation of levels of different systems are of the same form as their intra-subsystem level correlation. \footnote{This will be discussed further in the rest of Section \ref{sec2}}.
\item Extensiveness of the entropy is preserved, which is expected in a treatment considering weakly-interacting species.
\end{itemize}
While the first and third properties are immediate and expected, the second (and arguably the most important conclusion obtained from this treatment) reflects the fact that the system now behaves as if it consisted of $N$ indistinguishable particles of a single species with a joint spectrum of all the reaction subsystems (reactant, product, intermediates). This is also compatible with the constraints on the Lagrangian function being those of a single subsystem. 
This unique species, obeying the same spin-statistics as the original subsystems, arises naturally from the distribution and can be seen as the \textit{effective particle} of this distribution.
This picture also persists, a simpler form, upon inspecting the classical limit of the obtained distribution (Section \ref{sec3}). 
\footnote{In the herein derived statistical treatment, there is an implicit inclusion of the chemical equilibrium condition in the thermal equilibrium description. We effectively manage to replace an explicit chemical equilibrium treatment by a general thermal-statistical treatment.
This can only be achieved by taking advantage of the matching of the chemical potentials of the reacting subsystems in equilibrium
and associating a single chemical potential with the distribution. This chemical potential $\mu \equiv \mu(\{n_s^A\}, \{n_s^B\})$ is a function of (for a two subsystem reaction example) both $A$ and $B$ occupations, and fixed by the constraint equation $\sum_s n^A_s + \sum_{s'}n_{s'}^{B}= N$.}

We will refer to this thermal approach to equilibrium as the \textit{reaction thermalization}. 
A brief discussion on the conditions where this thermal reaction assumption is valid is included in Subsection \ref{subsec3.4}. The most important takeaway is that we require only that the occupations between the different subsystems can completely be described in a Boltzmann-thermal fashion. 

For this treatment to hold as a complete thermal and equilibrium description, we must handle only reaction networks \footnote{Or more generally, conversion networks which may not necessarily involve chemical reaction in its classical definition.} consisting of reactions of solely (true or apparent) unimolecular "steps" of the type $\ce{A <=> B <=> C ...}$, that have reached thermal and chemical equilibrium on all individual pathways. Reactants, products and long-lived intermediates with representation in the equilibrium must be thermalized and ergodic for this treatment to be physically sound. 
The existence of short-lived intermediates without expression at equilibrium (which may not obey the unimolecular "step" condition) do not affect the validity of the treatment.
The FD distribution ($+$) should be used for a system of interconverting open-shell atoms/molecules, and the BE picture for a conversion of closed-shell species.
\footnote{A full treatment, including correlations and fluctuations for all species in complex reaction networks, introduces additional subtleties, which are to be addressed in future work.}

For reaction networks consisting of more general reaction steps \footnote{This more general case will not be again derived as it trivially follows form the one-to-one case.} of the type $\ce{\nu_A A <=> \nu_B B}$ \footnote{Note that the reaction coefficients $\nu$ here are not under the definition of positive for products and negative for reactants, and take only positive values}, each individual molecule involved in a single step contributes with a replica of its spectrum to the total joint spectra, such that we the multiplicities become:
\begin{align}
& W^{BE}_{| \mtc{A}} = \prod_{\mtc{A}} \left[ \prod_{s \in \mathcal{A}}^{\mathcal{S}_{\mathcal{A}}} \frac{(g_s + n^{eq}_s - 1)!}{(g_s - 1)! n^{eq}_s!} \right]^{\nu_\mtc{A}} \quad  \notag \\
\\
& W^{FD}_{| \mtc{A}} = \prod_{\mtc{A}} \left[ \prod_{s \in \mathcal{A}}^{\mathcal{S}_{\mtc{A}}} \frac{g_s!}{(g_s - n^{eq}_s)! n^{eq}_s!} \right]^{\nu_\mtc{A}} \notag
\end{align}  
with $\mtc{A}$ summing over only the different species in the reaction network. 
Then, we have:
\begin{equation}
n^{\mathcal{A}, eq}_s = \nu_{\mtc{A}} \frac{g_s^{\mtc{A}}}{e^{\beta\varepsilon^{\mathcal{A}}_s - \beta\mu} \mp 1}
\end{equation}
naturally giving higher occupations to levels of the species having a higher reaction coefficient. Note that the reaction chemical potential $\mu$ is dependent on the reaction coefficients, since the joint spectra would now consist of $\nu_A$ replicas of the spectra of $A$ and $\nu_B$ replicas of the spectra of $B$ (for the example of a total of two species).

\subsection{Comparison with the independent subsystems approach to chemical equilibrium}
\label{subsec3.2}

\vspace{0.08cm} 

Contrary to the coupled formalism herein described, the approach traditionally adopted in the literature to statistically handle chemical equilibrium relies on the formalism of the previous section by treating the involved species as independent subsystems. This approach assumes \textit{a priori} static particle conversion dynamics. In doing so, it forces each subsystem's particle number to be fixed at values matching its the sum of its average occupations. This is valid for macroscopic systems under the scope of the thermodynamic limit, since effectively one does not need the particle number conservation constraint, as the averages are formally non-varying particle numbers and as such the system can be treated as being that of a grand canonical ensemble. 
However, one should know that this approach has limitations. It does not account for fluctuations that emerge in smaller (and not extremely large) systems. Additionally, it prevents one from using a canonical ensemble description.
Our approach differs in that it does not \textit{a priori} fix each systems average particle number, allowing them to fluctuate. Then this dominance or not of the average values is left as a consequence of the approximation, and not of any assumptions made.

Comparing the multiplicities of the two approaches, one can conclude, upon inspection, that $W_{| \mtc{A}} \geq W_{\times \mtc{A}}$, with the equality holding at the thermodynamic limit. 
The inequality results from the higher variance associated with $n^{eq}_s$ for it allows fluctuations over the collective spectrum, with a single constraint that the total number of entities sums to $N$; 
while the independent subsystems' multiplicity is calculated in relation to $n^{\mtc{A}}_s$, which only fluctuate within the spectra of the respective subsystem, and contains a particle number conservation constraint for each subsystem (thus being a system which is more constrained than the former case).
This might become even clearer if we consider that a conversion-coupled $A|B$ system under equilibrium as a independent $A\times B$ system with the extra freedom of the $\ce{A <=> B}$ transition, which 
increases the number of possibilities of distributing $N$ particles, giving rise to many more possible microstates and giving the coupled system a higher entropy.

If we are under conditions consistent with the thermodynamic limit, there is a formal "equivalence" of the two approaches. This is so since possible configurations that alter the instantaneous $N^A$, $N^B$ from their equilibrium values have negligible contribution when compared to the magnitude of their macroscopic, equilibrium values.
At this limit the coupled distribution formally reduces to that of independent subsystems and we can write the chemical potential as:
\eq{
\mu = \sum_{\mtc{A}}  \delta_{s \in \mathcal{A}} \ \nu_{\mtc{A}} \  \mu^{\mtc{A}} 
}
where $\delta_{s \in \mathcal{A}}$ is $1$ if $s \in \mtc{A}$, and $0$ otherwise. This way we have $\beta \mu \left( \sum\limits_{s \in \{\mtc{A}\}}^{\mathcal{S}} n^{eq}_s - N \right) \rightarrow \sum\limits_{\mtc{A}} \beta \mu^{\mtc{A}} \left( \sum\limits_{s \in \mtc{A}}^{\mathcal{S}_{\mtc{A}}} n^{eq}_s - \langle N^{\mtc{A}} \rangle \right) $ in Equation (\ref{eq6}), which effectively decouples the subsystems. 
This roughly corresponds, in the thermodynamic limit with an infinite number of levels, to an "equilibrium" chemical potential of the form of weighting of the chemical potentials of the subsystems, $\mu = \sum_{\mtc{A}} \frac{ \langle N^{\mtc{A}} \rangle }{N} \mu^{\mtc{A}}$.
Thus, as we approach the thermodynamic limit, the importance of intersubsystem correlations becomes increasingly small since the subsystems decouple. 
The reverse also holds - if we start moving away from the thermodynamic limit, the subsystems will start to behave in a correlated way due to the increasing importance of fluctuations. 
We can thus conclude that these two concepts are linked by the global particle conservation constraint.

It should be noted that the fluctuating character of certain systems described with our distribution may be incomplete. This is so since in building the variational procedure we assumed that the system is well described by average occupations. This may not hold for extreme cases where fluctuations are of utmost importance, such as when we're dealing of small systems from the thermodynamic point of view \cite{Hill1994}.

Information about the fluctuations of the occupation of a level belonging to a subsystem $A$ can be inferred from its variance:
\eq{
\langle \left( \Delta n^A_s \right)^2 \rangle = \langle \left( n^A_s \right)^2 \rangle - \langle n^A_s \rangle ^2
}
where $\langle n^A_s \rangle$ is what we have been referring to as simply $n_s^A$ or $n_s^{eq}$, the canonical average of the level occupation, respectively in the independent or conversion-coupled approach. For the coupled distribution this variance takes into account the configurations of the joint system:
\eq{
\langle n^{A}_s \rangle = \sum_{X} n_s^X  \frac{\exp \Big\{-\frac{E_X}{k_BT} \Big\}}{\sum_{X} \exp \Big\{-\frac{E_X}{k_BT} \Big\}}
}
$n_s^X$ being the occupation on configuration $X$, with system energy $E_X$. $X$ can be any of the configurations obtained by distributing $N$ particles among the joint set of levels, or, equivalently, a different configuration is obtained by exchanging particles not only between levels of the same subsystem but also of different subsystems. One should only sum over distinguishable configurations as per the indistinguishable identity of particles.
This same equation holds for the canonical average of the square of the occupation, replacing $n^X_s$ by $(n^X_s)^2$.
$N_A$ then also fluctuates, and its variance can be obtained from canonical averages on these same configuration occupations as:
\eq{
\langle \Delta N_A^2 \rangle =  \langle N_A^2 \rangle - \langle N_A \rangle ^2 =  \langle \left( \sum_{s \in A} n_s\right)^2 \rangle \ - \ \left( \sum_{s \in A} \langle n_s \rangle \right)^2
}

\subsection{The assumption of reaction thermalization}
\label{subsec3.4}
\vspace{0.08cm}

For the introduced thermalized description to hold within a single system (and not just as a canonical average), we also rely on the assumption of \textit{ergodic reaction equilibrium}: given sufficient time, the system will explore all accessible microstates, including those connected by crossing reaction pathways. \footnote{This justifies equating time averages with the ensemble averages used throughout this work.}
While ensemble averages can still be formally computed without ergodicity, the described correlations to exist in a single system requires the ergodicity assumption to hold. Without ergodicity, such correlations may only manifest partially or excluding some of the microstates. In particular, this can be true for systems with high activation barriers if not enough time is "supplied" to the system.
\footnote{In real, non-idealized systems, high barriers can hinder equilibrium or make it unclear whether equilibrium is achieved -  leading to an effective ergodicity breaking. As a result, fluctuations in quantities like particle number $\langle N^A \rangle$ may not match ensemble predictions within practical measurement times. This reflects a dependence on initial conditions of the specific system. }
Additionally to the ergodicity condition, for this form of correlation to be effective on the system we must assume that either: i) the system is in a long-lived equilibrium - effectively infinite if needed; or ii), that reaction activation barriers (if applicable) are not much larger than the thermal energy of the system ($\lesssim k_BT$), such that all microstates can be explored on reasonable timescales. 
\footnote{A realistic condition is a compromise between the two - the system's lifetime must increase as the height of (some of) the barriers increases compared to the thermal energy.} \footnote{Ergodicity within the available levels of each species does not necessarily imply reaction ergodicity, which also involves ergodicity along connecting states of different subsystems by transversing the reaction pathway.
Needless to say, if ergodicity is not realized within a subsystem, reaction ergodicity will immediately not hold as an assumption.}

It is important to note that the description of reaction dynamics of quantum gases should require quantum wave-field treatments of the involved species \cite{Zhang20231466, He2020}. 
The herein established framework does not require knowledge of the processes involved in conversion or of the reaction path that connects reactants and products. 
As such, in theory should still apply as long as the system is ergodic and thermalized - wether from the presence of a heat bath or a larger system which thermalizes the subsystems through 
entanglement phenomena \cite{Kaufman2016}.
Still, the specific processes involved in the conversion of species may hinder the thermalized description.
Another important point resides with the system maintaining its non/weakly-interacting nature in the presence of the interactions occurring during conversion. These will depend on the specific processes involved, and it is beyond the scope of this work to explore specific processes. 
However, even if such interactions are of a strong nature, if the lifetime of the system is large compared to the characteristic time of each reaction, the system should still behave consistently with that of a weakly-interacting gas.

\section{Classical limit}
\label{sec3}

We now analyze the classical (Maxwell-Boltzmann statistics) limit of the conversion-coupled distributions under chemical equilibrium.
The classical limit can be obtained by taking $n_s \ll 1 \ \ \forall \ s\in A,B$, such that quantum correlations arising from quantum statistics become negligible. This condition is roughly equivalent to the regime of a large (negative) chemical potential, $ -\mu \gg \varepsilon_s \ \forall s$. Under this approximation ${e^{\beta\varepsilon_s} e^{-\beta\mu} \mp 1} \approx e^{\beta\varepsilon_s} e^{-\beta\mu}$, and we can approximate Equation (\ref{eq7.1}) as:
\begin{equation}
\label{eqx.1}
n^{eq}_s \approx  N\frac{g_s e^{-\beta\varepsilon_s}}{\sum\limits_{\mtc{A}}\sum\limits_{s\in \mtc{A}} g_s e^{-\beta\varepsilon_s } } \equiv N \frac{g_s e^{-\beta\varepsilon_s}}{Z^{eq}_{MB}}
\end{equation}
resorting to the definition of the chemical potential $\ln \sum\limits_{s} g_s e^{\varepsilon_s} - \ln N = \beta\mu $. In Equation (\ref{eqx.1}) $Z^{eq}_{MB}= \sum_{\mathcal{A}} Z^{\mathcal{A}}_{MB} \equiv \sum\limits_{\mtc{A}} \sum\limits^{\mtc{S}_{\mtc{A}}}_{s \in \mtc{A}} g_s e^{-\beta\varepsilon_s} = \sum\limits_s^{\mtc{S}} g_s e^{-\beta\varepsilon_s}$ is the one-particle Maxwell-Boltzmann type partition function for the 
equilibrium system \footnote{In this section we focus only on the simpler form for one-to-one reactions. For general reaction coefficients as previously introduced we trivially obtain $Z^{eq}_{MB}= \sum_{\mathcal{A}} Z^{\mathcal{A}}_{MB} {}^{\nu_\mtc{A}}$}, which is formally a one-particle partition function for the composite set of levels. Since $Z_{MB}^{\mtc{A}} \equiv \sum\limits^{\mtc{S_A}}_{s \in \mtc{A}} g_s e^{-\beta\varepsilon_s}$ the one-particle conversion-coupled partition function is a sum of the independent subsystems one-particle partition functions.
We can then define \footnote{Dropping the MB subscript for simplicity}:
\eq{ \label{eqpA}
p^{\mtc{A}} \equiv Z^{\mtc{A}}\Big/ Z^{eq}
}
as the probability ($\leq 1$) that a random particle of the system will be found in a level belonging to the $\mtc{A}$ particle spectrum. Since the probability of the particle being found in level $s$ of the equilibrated composite system, $p_s = g_s^{\mtc{A}} \exp \{ -\ep^{\mtc{A}}_s \} / Z^{eq} = g_s^{\mtc{A}} \exp \{ -\ep^{\mtc{A}}_s \} /\  Z^{\mtc{A}} \times Z^{\mtc{A}} / \ Z^{eq}$, one has:
\eq{ \label{eqp}
p_s = p^{\mtc{A}} \ p_s^{\mtc{A}} 
}
with $p_s^{\mtc{A}} \equiv g^{\mtc{A}}_s\exp \{ -\ep_s \} / Z^{\mtc{A}}$, the probability that a particle restricted to subsystem $A$ is found in level $s$ (probability for $s$ in the independent subsystem case). 

We can arrive at the total system partition function in the classical limit and the canonical ensemble at the thermodynamic limit by using the grand canonical form of the partition function for the quantum systems as a starting point:
\eq{ \ln \Xi^{BE/FD} = \mp \sum_s \ln \left[ g_s\Big( 1\mp \exp \{\beta\mu -\beta\varepsilon_s\} \Big) \right]
}


The Helmholtz energy of the system, $F$ \footnote{sometimes simply referred to as the free energy} for this ensemble is given by $F=\Phi + \mu \langle N \rangle$, with  $\Phi= -\frac{1}{\beta}\ln \Xi$, the grand canonical potential. Since we work in the thermodynamic limit where the grand canonical ensemble average $\langle N \rangle$ particle number matches our conserved particle current value of $N$, we have the canonical ensemble partition function:
\begin{align}
\label{eqx.2} 
\ln \mathcal{Z}^{BE/FD}= -\beta F = \mp \sum_s &\ln \left[ g_s\Big( 1\mp \exp \{ \beta\mu -\beta\varepsilon_s\}  \Big)  \right] 
- \beta \mu N
\end{align}
We denote by $Z$ a one-particle partition function and $\mathcal{Z}$ a ensemble partition function.
In the classical limit of $n_s \ll  1 \ \forall 
 \ s\in \{ \mtc{A} \} = A,B,...$; as done for Equation (\ref{eqx.1}) and using $N \ln N - N \approx \ln N!$, we have our desired classical limit of the total system partition function given by:
\begin{equation}
\label{eqx.3}
\mathcal{Z}^{BE/FD} \approx \frac{{(Z^{eq}_{MB}})^N}{N!} \equiv \mtc{Z}^{clas}
\end{equation}
This partition function reinforces the conclusion from the quantum counterpart that the system behaves as an equilibrium ensemble of $N$ indistinguishable particles. \footnote{This form of a partition function agrees with the one previously derived by T. L. Hill for the case of two species in chemical equilibrium in a catalyzed reaction, making use of the multinomial theorem and the contributions of independent subsystem partition functions of all possible subsystem total occupations $(N^A, N^B)$ \cite{hillthermo}.} These effective particles thermally distribute among the joint spectrum,
following the standard indistinguishable Maxwell-Boltzmann statistics.
The form of the classical partition function of Equation (\ref{eqx.3}) also implicitly incorporates fluctuations of subsystem particle numbers, since states of different subsystems are weighted on equal footing with those belonging to the same subsystems.

One can easily compute variance of the occupation of each level belonging to a subsystem $A$ \footnote{Using $N_A \equiv N^\mtc{A}$ and $Z_{A} \equiv Z^{\mtc{A}}$ here such as not to clutter the notation}:
\begin{align} 
\label{eqfluc1}
\langle \left( \Delta n^{{A}}_s \right)^2 \rangle / N^2 = p_s^{{A}} - \left( p_s^{{A}} \right)^2 =& \ \frac{1}{Z_{eq}} g_s e^{-\ep_s}
- \frac{1}{Z_{eq}^2} g_s^2 e^{-2\ep_s} 
\end{align}
and of the subsystem particle numbers:
\eq{
\label{eqfluc2}
\langle \Delta N_{{A}}^2 \rangle / N^2 = p^{{A}} - \left( p^{{A}}\right) ^2 = Z_{{A}}/Z_{eq} - \left( Z_{{A}}/ Z_{eq} \right)^2
}

As an illustration of the incompleteness of the standard approaches for statistical chemical equilibrium, consider the classical case of the equilibrium partition function used for a system of many species $\{\mtc{A}\} = \{A,B,C...\}$ in chemical equilibrium. It is usually taken as \cite{levine2009physical}:
\begin{equation}
\label{eqy.2}
\mathcal{Z} = \prod_{\mathcal{A}} \frac{{(Z_{\mathcal{A}}})^{N^{\mathcal{A}}}}{N^{\mathcal{A}}!}
\end{equation}
where $Z_A$ is the Maxwell-Boltzmann one-particle partition function for species $\mathcal{A}$ and $N^\mtc{A}$ the fixed species particle number. This treatment is equivalent to mixing $N^A, N^B, N^C...$ independent particles of their respective species, without any chemical reactions occurring. 
This explains the label on this approach as a "frozen" or static approach - no reaction dynamics is being explicitly or implicitly taken into account, but only the form of the (static) equilibrium ensemble.
The equilibrium free energy in these approaches is of the type:
\eq{ \label{eqEX}
F^{eq}=\sum_{\mtc{A}} N^\mtc{A} \mu^\mtc{A}
}
One further has the species' chemical potential of the form:
\eq{ \label{muchem}
\mu^{\mtc{A}} = \left( \frac{\partial F^{eq}}{\partial N^\mtc{A}} \right)_{T,V} =  -\frac{1}{\beta} \frac{\partial \ln \mathcal{Z}}{\partial N^\mtc{A}} = \mu_\mtc{A}^{(d)} + k_B T \ln N^{\mtc{A}}
}
where $\mu_{\mtc{A}}^{(d)} \equiv -\frac{1}{\beta} \ln Z^{\mtc{A}}$ is the chemical potential for an ensemble of ideal distinguishable particles of $\mtc{A}$, and differs from the actual chemical potential of particles of the ensemble because of their indistinguishable nature, translated in the factor $k_B T \ln N^{\mtc{A}}$.
This then results in the traditional law of mass action for the simple example reaction \ce{A <=> B}, by making use of the fact that $\mu^A=\mu^B$:
\eq{ \label{massac}
Z^B/Z^A = N^B/N^A
}
This type of result is valid exactly only for infinitely large system at the thermodynamic limit, since for more realistic systems fluctuations become important and we can find the system on non-equilibrium composition states. The entropic effects of these fluctuations are also neglected in this treatment. \footnote{The framework presented in this work does not have a chemical equilibrium equality condition and as such has no associated law of mass action. It doesn't need one, for the calculation of the set of $\{N^\mtc{A}\}$ is direct if one has access to the set of one-particle partition functions of the system $\{Z^\mtc{A}\}$, via $N^\mtc{A} = p^\mtc{A} N$}

Compare the values of the level $s \in A$ occupation variances and composition fluctuation variances ($\langle \Delta N_{{A}}^2 \rangle$) obtained from the independent subsystem formulation:
\begin{align}
& \langle \left( \Delta n^{{A}}_s \right)^2 \rangle / N_A^2 = \frac{1}{Z_{A}}g_s e^{-\ep_s} - \frac{1}{Z_{A}^2} g_s^2 e^{-2\ep_s}, \\
& \langle \Delta N_{{A}}^2 \rangle=0 \notag
\end{align}
with the composition fluctuation variances Equation (\ref{eqfluc1}) and (\ref{eqfluc2}). It can be inferred from the relation between $Z_{eq}$ and $Z_{A}$ in Equation (\ref{eqpA}) (which also trivially holds as a relation between $\langle N_A \rangle$ and $N$) that for the coupled interconverting subsystems:
\eq{
\langle \left( \Delta n^{{A}}_s \right)^2 \rangle / \langle N_A \rangle ^2 =  \frac{1}{p^A Z_A} g_s e^{-\ep_s}/Z_{A} - \frac{1}{Z_{A}^2}g_s^2 e^{-2\ep_s}
}
for $s \in A$, and thus (if we take $\langle N_A \rangle = N_A$, which is an arbitrary choice of the independent subsystem $A$ particle number) we see that the per-particle variances of occupations and subsystem particle number are greater or equal (equal when $p_A = 1$) for coupled reacting subsystems when compared to the independent subsystems. 
This is expected following the discussion in Subsection \ref{subsec3.2}.

\section{Conclusions} 
\label{conc}

In this work we demonstrated how a system of weakly-interacting, interconverting quantum gases of atoms and molecules in thermal and chemical equilibrium can be described using a quantum distribution over a collective spectrum
of the reacting species. This relinquishes the need for the introduction of the chemical equilibrium condition - which now arises from a more complete thermal equilibrium condition under a global particle number conservation. 
As a result, composition fluctuations are intrinsic to the model and are naturally included as distinct microstates in the ensemble description. The entropy of this mixture is increased when compared to the case where no interconversion of species is present.

The derived distribution predicts effective Bose-Einstein and Fermi-Dirac correlations between the energy states of different species coupled by total particle conservation. 
These correlations hold within a single system under time evolution as long as the system, including on the interconversion pathways, is ergodic.

Although our theory is firmly grounded in quantum statistical mechanics, it applies strictly to canonical Gibbs states of non-interacting or weakly-interacting particles - systems without coherence that remain in full thermal equilibrium. The species involved are assumed to remain non‑interacting at the macroscopic level despite the possibility of important interactions during the reaction process. 

Notwithstanding the fact that the simultaneous conditions of thermalization and quantum behavior may be hard to fulfill in real systems, quantum gases provide a robust abstraction, since any approximation employed to describe actual systems can naturally be obtained from this more general description. 
One special case that can be obtained this way is the classical limit of the ideal gas. In this limit our formulation expands upon traditional chemical equilibrium by accounting for compositional fluctuations that traditional models typically neglect. 

This work represents an initial, simplified realization of a potentially broader formalism. We have focused solely on one-reactant-one-product reactions, but the framework is in principle extendable to any kind of complex reaction networks. 

Finally, we reinforce how the concept of intrinsically coupled energy spectra arising solely from chemical equilibrium may offer profound new insights into how microscopic quantum correlations shape macroscopic thermodynamic behavior. We believe this emerging perspective warrants further theoretical and experimental exploration.

\section*{Acknowledgments}

The author is supported by national funds via the Portuguese Foundation for Science and Technology (FCT).





\end{document}